\documentclass[aps,pra,twocolumn]{revtex4}
\usepackage{graphicx}
\usepackage{amsmath}
\usepackage{amssymb}

\def\lsim{\
  \lower-1.2pt\vbox{\hbox{\rlap{$<$}\lower5pt\vbox{\hbox{$\sim$}}}}\ }
\def\gsim{\
  \lower-1.2pt\vbox{\hbox{\rlap{$>$}\lower5pt\vbox{\hbox{$\sim$}}}}\ }

\begin{document}

\title{Are superfluidity and superconductivity related \\to the spontaneous breaking of $U(1)$ symmetry?}

\author{Maksim Tomchenko}
\affiliation{Bogoliubov Institute for Theoretical Physics, 14b,
Metrolohichna Str., Kyiv 03143, Ukraine}

\date{\today}
\begin{abstract}
It is widely believed that superfluidity and superconductivity are
caused by the spontaneous breaking of global $U(1)$ symmetry. We
investigate this question for quantum-mechanical systems using the
following precise criterion for spontaneous symmetry breaking (SSB):
the system's Hamiltonian  and boundary conditions are invariant
under transformations of a certain symmetry group, while the ground
state is not. Through a simple analysis, we show that in real-world
systems, which are always finite, spontaneous breaking of $U(1)$
symmetry does not occur. Therefore, superfluidity and
superconductivity can be related to SSB only in infinite systems,
which do not exist in our Universe. We also discuss the relationship
between the theory of SSB and reality.
\end{abstract}

\maketitle

\textbf{Keywords:} superfluidity, superconductivity, $U(1)$
symmetry, spontaneous symmetry breaking

\section{Introduction}
It is often claimed in the literature that superfluidity and
superconductivity are related to the spontaneous breaking of global
$U(1)$ symmetry; see the reviews
\cite{bogquasi2,grib1970,kirzhnits1978,anderson1984,miransky1994,vollhardt2000,nambu2009,shirkov2009,pekker2015,wezel2019,shimano2020,powell2020}.
A detailed analysis of this problem has been performed in
Refs.~\cite{bogquasi2,grib1970,miransky1994,u1sf}. Since
superconductivity is the superfluidity of electron pairs, below we
consider only superfluidity. From a physical point of view,
superfluidity is caused by a condensate of particles ($^4$He atoms,
pairs of $^3$He atoms, or electron pairs in superconductors). At the
same time, theory says that if spontaneous breaking of $U(1)$
symmetry occurs, it is precisely this breaking that leads to the
formation of such a condensate
\cite{bogquasi2,kirzhnits1978,miransky1994}.

The works
\cite{bogquasi2,grib1970,kirzhnits1978,anderson1984,miransky1994,vollhardt2000,nambu2009,shirkov2009,pekker2015,wezel2019,shimano2020,powell2020}
consider infinite systems. However, in nature all systems are
finite. In the literature,  the distinction between finite and
infinite systems is usually not made: the thermodynamic limit ($N, V
\rightarrow \infty, N/V=\mathrm{const}$) is used,  and it is
implicitly assumed that passing to this limit does not alter the
properties of the system. It is known, however, that certain
properties change upon passing to the thermodynamic limit: these
include Bose condensation in one dimension (1D) and two dimensions
(2D), crystallization in 1D and 2D, and  phase transitions in any
dimension \cite{mainwood2006}. It has also been shown that
spontaneous breaking of $U(1)$ symmetry is absent in a finite system
but occurs in an infinite one for the following systems: an ideal
gas of spinless bosons \cite{miransky1994}, a system of interacting
spinless bosons \cite{u1sf}, and a BCS superconductor
\cite{greiter2005}. The analysis in these three papers is based on
the derivation of the second-quantized ground-state wave function
$|0\rangle$ of the system. Below, we will see that the question
posed in the title can be answered in a much simpler way.

\section{Superfluidity vs. spontaneous symmetry breaking}

To answer whether spontaneous symmetry breaking (SSB) is present, we
need an exact mathematical criterion for SSB and an accurate method
to verify it. For a finite system, there are two strict definitions
of SSB: quantum~\cite{u1sf, sgs} and statistical~\cite{bogquasi2}.
Quantum definition: \emph{The boundary-value problem (the
Hamiltonian and boundary conditions) is invariant under a certain
symmetry, whereas the ground state is not.} The statistical
definition operates with the equilibrium state instead of the ground
state. We will rely on the quantum definition because calculations
based on it are usually simpler, and the quantum definition is in
fact a special case ($T=0$) of the statistical one. The quantum
criterion for SSB in an infinite system does not involve boundary
conditions: the Hamiltonian is invariant, but the ground state is
not. This definition is commonly accepted in quantum field theory
\cite{miransky1994}. Some authors also use this criterion for finite
systems. In nature, however, the symmetry of the Hamiltonian is
often broken by boundary conditions, with the result that the ground
state also loses its invariance. This is obviously not SSB.
Therefore, we will use the aforementioned quantum criterion, which
takes boundary conditions into account.

Since in non-relativistic systems particles are neither created nor
annihilated, in a finite quantum-mechanical system the number of
particles is conserved and definite (in this case, ``particle''
refers to an atom for electrically neutral superfluid systems and an
electron for charged  systems). Therefore, the ground-state wave
function $|0\rangle$ must be an eigenfunction of the particle number
operator $\hat{N}$,
\begin{equation}
\hat{N}|0\rangle=N|0\rangle  \label{1}%
\end{equation}
(recall that
$\hat{N}=\sum_{\sigma}\sum_{\mathbf{k}}\hat{a}_{\mathbf{k},\sigma}^{+}\hat{a}_{\mathbf{k},\sigma}$,
where $\hat{a}_{\mathbf{k},\sigma}^{+}$ and
$\hat{a}_{\mathbf{k},\sigma}$  are the creation and annihilation
operators for a particle with momentum $\mathbf{k}$, and $\sigma$
denotes the discrete spin projections of a particle), and the
Hamiltonian of the system should commute with $\hat{N}$:
$[\hat{H},\hat{N}]=0$. The latter always holds for the exact
Hamiltonian
\begin{align}
\hat{H}  &  =-\frac{\hbar^{2}}{2m}\sum\limits_{\sigma}\int\limits_{V}d\mathbf{r}\hat{\psi}%
^{+}(\mathbf{r},\sigma,t)\Delta \hat{\psi}(\mathbf{r},\sigma,t)\nonumber\\
&  +\frac{1}{2}\sum\limits_{\sigma,\sigma^{\prime}}\int\limits_{V}d\mathbf{r}d\mathbf{r}^{\prime}U(|\mathbf{r}%
-\mathbf{r}^{\prime}|)\hat{\psi}^{+}(\mathbf{r},\sigma,t)\hat{\psi}^{+}%
(\mathbf{r}^{\prime},\sigma^{\prime},t)\nonumber\\
& \times\hat{\psi}(\mathbf{r}^{\prime
},\sigma^{\prime},t)\hat{\psi}(\mathbf{r},\sigma,t) \label{0}%
\end{align}
of a system of $N$ indistinguishable interacting bosons or fermions.
Therefore, such a Hamiltonian is invariant under the $U(1)$
transformation defined by the operator $\hat{U}_{\varphi}=
e^{i\varphi\hat{N}}$:
\begin{equation}
\hat{\psi}(\mathbf{r},\sigma,t)\rightarrow\hat{U}_{\varphi}^{-1}\hat{\psi}(\mathbf{r},\sigma,t)\hat{U}_{\varphi}
=e^{i\varphi}\hat{\psi}(\mathbf{r},\sigma,t),  \label{0-1}%
\end{equation}
\begin{equation}
\hat{H}\rightarrow\hat{U}_{\varphi}^{-1}\hat{H}\hat{U}_{\varphi}=\hat{H}.  \label{0-2}%
\end{equation}
Given property (\ref{1}), the ground-state wave function $|0\rangle$
is also invariant under the $U(1)$ transformation:
\begin{equation}
|0\rangle\rightarrow\hat{U}_{\varphi}|0\rangle\equiv
e^{i\varphi\hat{N}}|0\rangle
=e^{i\varphi N}|0\rangle.  \label{2}%
\end{equation}
Consequently, the ground state is non-degenerate with respect to
$U(1)$ symmetry. If the ground state is degenerate and corresponds
to several wave functions, then property (\ref{1}) must hold for
each of them, as the number of particles is fixed. Therefore, even
if the ground state is degenerate, this degeneracy is not related to
$U(1)$ symmetry. Note that degeneracy of the ground state is
possible if the conditions of the Courant--Hilbert theorem
\cite{gilbert} are violated; the violating factors include, among
others, the particle's spin and the infinite size of the system; for
more details, see Ref.~\cite{sgs}. According to Kramers' theorem
\cite{kramers1930}, any state of a system with half-integer total
spin is at least doubly degenerate if the system's Hamiltonian is
invariant under time reversal.

In real systems, there is usually  weak diffusion of particles
across the boundary (so that the total number of particles
fluctuates very slightly). Mathematically, this means that the
boundary conditions for $\hat{\psi}(\mathbf{r},t)$ are not strictly
invariant under $U(1)$ transformations. However, this effect can be
neglected, since it does not affect bulk properties, including
superfluidity. Therefore, such boundary conditions can be considered
invariant under $U(1)$ rotations. Periodic and zero boundary
conditions possess this latter property.

Thus, for a finite system, the Hamiltonian, boundary conditions, and
ground state are invariant under $U(1)$ transformations. This means,
according to our criterion, that spontaneous breaking of global
$U(1)$ symmetry is absent.

Equations (\ref{1}) and (\ref{2}) are universal and hold for all
finite-size many-particle systems in which the number of particles
is conserved:  quantum gases and quantum liquids ($^{3}$He and
$^{4}$He),  superconductors, and all other systems except plasmas.
Therefore, \emph{in such systems, spontaneous breaking of global
$U(1)$ symmetry does not occur}.

Note that for a superfluid Bose gas and He II, an \emph{exact}
second-quantized ground-state wave function $|0\rangle$ has been
found \cite{u1sf}. It satisfies properties (\ref{1}) and (\ref{2}).
For more complex systems, it may be impossible to write such a
function in explicit form. However, to determine whether spontaneous
$U(1)$ symmetry breaking is present, formulas (\ref{1}) and
(\ref{2}) are sufficient.

For an infinite system, the situation is more complicated, since
infinity possesses paradoxical properties \cite{u1sf}, and the very
definition of SSB for an infinite system is still a matter of debate
\cite{baker2013}. In particular, the ground state of an infinite
Bose gas or an infinite superconductor is infinitely degenerate,
since for this state \cite{miransky1994}
\begin{equation}
\hat{U}_{\varphi}|\theta\rangle=|\theta+\varphi\rangle.\label{44}%
\end{equation}
According to the analysis in Ref.~\cite{u1sf},  this degeneracy is
related to the indeterminacy of the number of particles in an
infinite system: if $N=\infty$, then $N=N+1=N+2=\ldots$ and
$E_{0}(N)=E_{0}(N+1)=E_{0}(N+2)=\ldots$. At the same time, it is
easy to see that all such systems with different values of $N$ are
one and the same system; that is, degeneracy is absent. This
duality---infinite degeneracy and, at the same time,
non-degeneracy---is a manifestation of the strange properties of
infinity.

Many authors associate superfluidity with SSB because of the
following quantum-field-theory formula for a condensate of atoms:
\begin{equation}
\langle0|\hat{\Psi}(\mathbf{r},t)|0\rangle=\Psi(\mathbf{r},t).
\label{3} \end{equation} %
If this expectation value is nonzero, we have degeneracy of the
ground state with respect to $U(1)$ symmetry, and thus spontaneous
breaking of $U(1)$ symmetry \cite{miransky1994,u1sf}. Since an
atomic condensate implies superfluidity (if the Landau criterion is
satisfied), Eq.~(\ref{3}) points  to a connection between
superfluidity and  SSB. Let us consider this issue. A strict and
universal quantum-mechanical definition of an atomic condensate is
based on the diagonalized expansion of the single-particle density
matrix \cite{leggett2006},
\begin{equation}
F_{1}(\mathbf{r},\mathbf{r}^{\prime},t)  =
\sum\limits_{j=1}^{\infty}\lambda_{j}(t)\phi^{*}_{j}(\mathbf{r}^{\prime},t)\phi_{j}(\mathbf{r},t),
     \label{4} \end{equation}
where the natural orbitals $\phi_{j}(\mathbf{r},t)$ form a complete
set of orthonormal functions,  $\lambda_{j}$ are the occupation
numbers of the single-particle states $\phi_{j}(\mathbf{r},t)$, and
$\lambda_{1}+\ldots +\lambda_{\infty} =N$. State $1$ is
macroscopically occupied if $\lambda_{1} \sim N$. All other
definitions of a condensate are either special cases of this one or
incorrect. For example, the introduction of a condensate via
off-diagonal long-range order (ODLRO),
\begin{equation}
\lim_{|\mathbf{r}-\mathbf{r}^{\prime}|\rightarrow
\infty}{F_{1}(\mathbf{r},\mathbf{r}^{\prime},t)}  =\Psi(\mathbf{r},t)\Psi^{*}(\mathbf{r}^{\prime},t),\label{6}%
\end{equation}
is a special case of definition (\ref{4}), see
Ref.~\cite{leggett2006}. If the system is infinite, then approaches
(\ref{3}) and (\ref{4}) are usually (though not always
\cite{sarasua2012}) equivalent and lead to the same condensate. For
a finite system, the approach based on Eq.~(\ref{4}) always works
and sometimes yields a nonzero condensate, whereas formula (\ref{3})
always yields zero (according to the rules of second quantization,
since the number of particles is fixed in a finite system). Given
these properties, for a quantum-mechanical system, determining the
condensate based on formula (\ref{3}) is unreliable, and for a
finite system, it is simply incorrect (however, in quantum field
theory, where the number of particles is indeterminate due to
passing to the thermodynamic limit and the possibility of particle
creation and annihilation, applying formula (\ref{3}) is entirely
justified).

Thus, our analysis clearly shows that in finite non-relativistic
superfluid systems, including superconductors, spontaneous breaking
of $U(1)$ symmetry does not occur. Such symmetry breaking is
possible only in hypothetical infinite systems, which do not exist
in nature. Therefore, \emph{in real, always finite systems,
superfluidity is caused by a condensate of atoms (or pairs of atoms,
or pairs of electrons) and is in no way related to spontaneous
breaking of $U(1)$ symmetry}. Thus, the common assertion that
superfluidity and superconductivity are caused by spontaneous
breaking of $U(1)$ symmetry is incorrect unless it is accompanied by
the remark that we are dealing with an infinite system from another
universe. This point is important for understanding the nature of
superfluidity and superconductivity.

Of course, our conclusion depends on the SSB criterion. If we choose
a different criterion, we may arrive at a different conclusion.
However, other criteria are less strict, so a different conclusion
would most likely be simply a consequence of the criterion's lack of
strictness.

Note that virtually all of modern physics at the very cutting edge
is ``physics of other worlds'' (such as modifications of special and
general relativity, quantum theories of gravity, cosmology, etc.).
However, physicists studying condensed matter can afford to
distinguish between our world and another world \cite{note1}.

Articles
\cite{curie1894,earman2004,landsman2013,fraser2016,wallace2018}
discuss the relationship between theories of SSB and reality. The
main problem here is that the ground state of a finite
quantum-mechanical many-body system is either non-degenerate or
finitely degenerate \cite{note2}, so only the spontaneous breaking
of \emph{discrete} symmetry is possible. Infinite degeneracy and
spontaneous breaking of \emph{continuous} symmetry are possible in
theory only for an infinite system. However, in nature all systems
are finite, yet nature seems to exhibit spontaneous breaking of
continuous symmetries. In our view, Curie's principle provides a
solution to this paradox, at least for condensed matter: ``When
certain effects show a certain asymmetry, this asymmetry must be
found in the causes which gave rise to it''
\cite{earman2004,curie1894}. Indeed, the paradox arises because the
theory fails to account for weak external influences, which are
always present in nature, and because of an inappropriate definition
of SSB for a finite system (SSB in finite and infinite systems is
usually defined identically: the system's Hamiltonian is invariant,
while the ground state is not). If we take into account external
perturbations in the Hamiltonian and boundary conditions in the
definition of SSB, the paradox disappears: as a matter of fact, in
nature, SSB \emph{does not} occur, since the symmetry of the
Hamiltonian is broken by boundary conditions and/or external fields
\cite{sgs}. This holds true for the liquid--crystal transition
\cite{sgs} and apparently for all other transitions (interestingly,
if the Hamiltonian and boundary conditions are invariant under the
rotation group $SO(3)$, then the ground state of the crystal is also
invariant; such a crystal does not exhibit Bragg peaks \cite{sgs}).
The ordering of the new phase is determined by the internal
properties of the system, but the impetus for the transition to the
new phase is provided by external forces; in this sense, symmetry
breaking in nature can be considered induced. Thus, in nature, SSB
does not occur: the symmetry of the ground state is lower than that
of the Hamiltonian due to the weak influence of an external field
(explicit symmetry breaking) and/or the influence of boundaries
(also explicit symmetry breaking, according to our criterion). We
assume that this conclusion holds for any continuous symmetry,
whether spatial or internal.

In recent years, the concept of generalized global symmetries has
been developing intensively; it extends the traditional notion of
symmetry and plays an important role in quantum field theory and the
description of topological phases of matter
\cite{mcgreevy2023,chen2025}. This field includes invertible
higher-form symmetries \cite{gaiotto2015,cordova2022,bhardwaj2023},
invertible higher-group symmetries \cite{cordova2022,bhardwaj2023},
non-invertible symmetries \cite{cordova2022,bhardwaj2025}
(categories instead of groups), as well as subsystem symmetries
\cite{cordova2022,distler2022}. For invertible generalized
symmetries, including group realizations of subsystem symmetries,
our definition of SSB is naturally generalized by replacing the
usual transformations with the corresponding generalized symmetry
operations. However, the definition of SSB must be significantly
modified for non-invertible symmetries, where some symmetry
operators do not have inverses. In this article, we consider only
ordinary symmetries.

\section{Conclusion}
Based on the quantum definition of SSB, we have shown that
superfluidity in real-world (always finite) quantum-mechanical
systems is not related to the spontaneous breaking of $U(1)$
symmetry. However, it is commonly believed in the literature that
superfluidity (including superconductivity) in real systems is
caused precisely by such a symmetry breaking. This misleading view
has penetrated  encyclopedias, including Wikipedia, and even
YouTube. Yet in fact, superfluidity is related to the spontaneous
breakdown of $U(1)$ symmetry only for an idealized, infinite system.
Therefore, it is important to clearly distinguish between  ideal
theoretical SSB and  natural symmetry breaking. The latter occurs in
finite systems and is explicit (according to our criterion),
although it appear spontaneous. Such symmetry breaking can be called
induced SSB, because it is induced by external factors (including
boundaries). This important point has been little discussed in the
literature, resulting in a certain disconnect between theory and
reality.

\begin{acknowledgments}
The author thanks S. Sharapov for drawing his attention to
Ref.~\cite{kats1984}, which gave impetus to this research. This work
was supported by the National Academy of Sciences of Ukraine and the
Ministry of Education and Science of Ukraine.

AI tools, Google Gemini (4 Hydrogen) and OpenAI ChatGPT (GPT-6),
were used to assist with literature searches, discussions of
theoretical concepts, and English-language checking.
\end{acknowledgments}


\begin{thebibliography}{99}
\bibitem {note1} After winning the Nobel Prize, of course, one no longer needs to
make this distinction; but until then, it is better to do so.
\bibitem {note2} To our knowledge, this statement has not been strictly
proven and follows from well-known quantum-mechanical solutions. The
spin degeneracy of the ground state is equal to $2S+1$, where $S$ is
the total spin of the system. The degeneracy of the spatial part of
the ground-state wave function should not be infinite, since this
would require the presence, among the ground-state wave functions,
of those with an infinite number of nodes; however, a large number
of nodes typically implies a large gradient of the wave function
over a significant region of space, which makes such a state
energetically high, a feature incompatible with the ground state.
\end{thebibliography}
\end{document}